\documentclass[showpacs,twocolumn,aps,prl]{revtex4}
\usepackage{amssymb}
\usepackage{amsmath}
\usepackage{epsfig}
\begin{document}

\title{Effective Field Theory Approach to Gravitationally Induced Decoherence}

\author{M. P. Blencowe}
\email{miles.p.blencowe@dartmouth.edu}
\affiliation{Department of Physics and Astronomy, Dartmouth College, Hanover, New Hampshire 03755, USA}
\date{\today}

\begin{abstract}
Adopting the viewpoint that the standard perturbative quantization of general relativity provides an effective description of quantum gravity that is valid at ordinary energies, we show that gravity as an environment induces the rapid decoherence of stationary matter superposition states when the energy differences in the superposition exceed the Planck energy scale.
\end{abstract}

\pacs{04.60.Bc, 03.65.Ta, 03.65.Yz}

\maketitle

\textit{Introduction}.--- The emergence of the macroscopic classical world from the microscopic quantum world is commonly understood to be a consequence of the fact that any given quantum system is open, unavoidably interacting with unobserved environmental degrees of freedom that will  cause initial quantum superposition states of the system to decohere, resulting in classical mixtures of  either/or  alternatives~\cite{caldeira83,joos85,zurek91}. Consider, for example, a system consisting of a vibrating micrometer scale silicon wire in ultrahigh vacuum at dilution fridge temperatures  ($\sim 10~\mathrm{mK}$). Assuming a realizable quality factor $Q\sim 10^5$ that is limited by clamping radiation loss~\cite{cole11} and elastic strain-coupled two level system defects within the wire~\cite{remus09}, an initial center of mass coherent state superposition with separation $\Delta x\sim 1~{\mathrm{nm}}$ will decohere in about a picosecond, rapidly enforcing classicality in the dynamics of the vibrating wire.  Suppose, however, that the common sources of decoherence are removed through levitating the silicon mass by optical~\cite{li11,romero11a} or other means~\cite{romero12}. Can the coherence times of center of mass superposition states be increased without bound by removing the effects of clamping and defect loss in this way and minimizing the interaction with the electromagnetic environment? More generally, can systems of arbitrarily increasing mass/energy be placed in non-classical states, such as center of mass  quantum superposition states?  

Gravity has been invoked in various ways as playing a possible fundamental role in enforcing classicality of matter systems beyond a certain scale~\cite{feynman03,karolyhazy86,diosi87,ellis89,penrose89,ghirardi90,milburn91,blencowe91,stodolsky96,pearle96,anastopoulos96,kay98,anandan99,power00,amelino00,lamine06,gambini07,adler07,anastopoulos07,anastopoulos08,breuer09,gambini10,romero11b}. Certainly, one environment that cannot be avoided is the stochastic gravitational radiation background arising from the Big Bang and other sources~\cite{allen88,lamine06}. A clue as to the possible effect this environment might have on a low energy quantum matter system comes from the fact that the space-time metric in Einstein's equations couples to the system via its energy-momentum tensor. For a stationary system, only its rest energy should be relevant for the decoherence dynamics of an initial quantum superposition state. 
Consider for the moment a model oscillator system coupled via its energy to an oscillator environment, described by the Hamiltonian
\begin{equation}
H=\hbar\omega_0 a^{\dagger}a\left(1+\sum_i \lambda_i \frac{q_i}{\Delta_i}\right) +\sum_i \left(\frac{p_i^2}{2m_i}+\frac{1}{2}m_i\omega_i^2 q_i^2\right),
\label{modelhamiltonianeq}
\end{equation}                 
where $\omega_0$ is the system oscillator's frequency and  $\Delta_i=\sqrt{\hbar/(2m_i\omega_i)}$ is the $i$th bath oscillator's zero-point uncertainty. Assuming an Ohmic bath spectral density $J(\omega)/(\hbar\omega_0)^2=\pi \sum_i\lambda_i^2 \delta(\omega-\omega_i)=C \omega/\omega^2_0$, for  weak system-bath dimensionless coupling $C$ in the high temperature limit we obtain the following time evolution of  the system density matrix in the Born-Markov approximation:
\begin{equation}
\rho_{n \tilde{n}}(t)=e^{-C(k_B T/\hbar)  (n-\tilde{n})^2 t}\rho_{n \tilde{n}}(0),
\label{evolutioneq}
\end{equation}
where $T$ is the oscillator bath temperature.
Notice that the thermal oscillator environment induces decoherence without damping: initial superpositions of different Fock states $|n\rangle, |\tilde{n}\rangle$ decay into mixtures of these states. In other words, the environment ``localizes" the system energy. By analogy, and with the aid of dimensional analysis, we might therefore expect that a stochastic gravitational environment will similarly decohere a matter system initially in a superposition of say two different rest  energy states $E$ and $\tilde{E}$ with a rate given by
\begin{equation}
\Gamma_{\mathrm{decohere}}\sim\frac{k_{\mathrm{B}} T}{\hbar}\left(\frac{E-\tilde{E}}{E_{\mathrm{P}}}\right)^2,
\label{decohereeq}
\end{equation} 
where  $E_{\mathrm{P}}=\sqrt{\hbar c^5/G}$ is the Planck energy and we assume for simplicity a thermal graviton environment at temperature $T$. 

In the following, we shall derive Eq.~(\ref{decohereeq})--including the missing dimensionless numerical factor--by applying standard perturbative quantum field theory techniques to gravity~\cite{donoghue94,campos98,arteaga04}. The justification for such an approach follows from the fact that we are considering laboratory scale systems, where the matter is localized to regions of small curvature. As with other low energy effects, such as the quantum gravity correction to the Newtonian potential between two ordinary masses~\cite{donoghue94}, it should be possible to {\emph{quantitatively}} evaluate gravitationally induced decoherence rates by employing standard perturbative quantum gravity as an {\emph{effective}} field theory~\cite{donoghue94,burgess04}; whatever the final form the eventual correct quantum theory of gravity takes, it must converge in its predictions with the effective field theory description at low energies.         

\textit{Effective field theory derivation}.--- In order to be able to construct matter system states starting from a generally covariant field theory, we will adopt as a simple model system a massive scalar field $\phi(x)$ with mass parameter $m$ corresponding to that of a nucleon. Expanding the Einstein-Hilbert action to second order in metric deviations from Minkowski space-time, $g_{\mu\nu}=\eta_{\mu\nu}+\kappa h_{\mu\nu}$, we have:
\begin{equation}
S[h_{\mu\nu},\phi]\approx S_{{S}}[\phi]+S_{{E}}[h_{\mu\nu}]+S_{{I}}[h_{\mathrm{\mu\nu}},\phi],
\label{systemenvironmenteq}
\end{equation}
where $\kappa=\sqrt{32\pi G}$ (from now on we for the most part use natural units with $\hbar=c=1$), and the system, environment, and interaction actions are respectively:
\begin{equation}
S_{{S}}=-\frac{1}{2}\int d^4 x (\eta^{\mu\nu}\partial_{\mu}\phi\partial_{\nu}\phi +m^2 \phi^2)
\label{syseq}
\end{equation}
\begin{eqnarray}
S_{{E}}&=&\int d^4 x \left(-\frac{1}{2}\partial^{\rho}h^{\mu\nu}\partial_{\rho}h_{\mu\nu}+\partial_{\nu}h^{\mu\nu}\partial^{\rho}h_{\mu\rho}\right.\cr
&&\left.-\partial_{\mu}h\partial_{\nu}h^{\mu\nu}+\frac{1}{2}\partial^{\mu}h\partial_{\mu}h\right)
\label{enveq}
\end{eqnarray}
\begin{equation}
S_{{I}}=\int d^4 x\left(\frac{\kappa}{2} T^{\mu\nu}(\phi)h_{\mu\nu}+ \frac{\kappa^2}{4}U^{\mu\nu\rho\sigma}(\phi)h_{\mu\nu}h_{\rho\sigma}\right),
\label{inteq}
\end{equation}
where $T_{\mu\nu}(\phi)=\partial_{\mu}\phi\partial_{\nu}\phi-\frac{1}{2}\eta_{\mu\nu}\partial_{\rho}\phi\partial^{\rho}\phi-\frac{1}{2}\eta_{\mu\nu}m^2\phi^2$ is the scalar field energy-momentum tensor and the explicit form of the quadratic in $\phi$ tensor $U_{\mu\nu\rho\sigma}(\phi)$ can be found in Ref.~\cite{arteaga04}.

The closed time path integral approach~\cite{calzetta08} gives the following formal expression for the scalar matter system density matrix:
\begin{eqnarray}
&&\rho_S[\phi,\phi',t]=\int d\phi_0d\phi'_0 \rho_S [\phi_0,\phi_0',0]\cr
&&\times\int_{\phi_0}^{\phi}[d\phi^+]\int_{\phi_0'}^{\phi'}[d\phi^-]e^{\{i(S_S[\phi_+]-S_S[\phi^-]+S_{IF}[\phi^+,\phi^-])\}},\cr
&&\label{ctpeq}
\end{eqnarray}
where $S_{IF}$ is the Feynman-Vernon influence action that gives the effect of the thermal graviton environment on the scalar matter system. Evaluating $S_{IF}$ to lowest, quadratic order in $\kappa$ with harmonic gauge fixing term inserted in $S_E$, we obtain from Eq.~(\ref{ctpeq}) the following Born-approximated master equation for the scalar system:
\begin{eqnarray}
{\partial_t\rho_S(t)}&=&-i[H_S,\rho_S(t)]-\int_0^t d\tau\int d{\bf{r}}d{\bf{r}'}\Bigl\{ N({\bf{r}}-{\bf{r}'},\tau)\cr
&&\times\Bigl(2[T_{\mu\nu}({\bf{r}}),[T^{\mu\nu}({\bf{r}'},-\tau),\rho_S(t)]]\cr
&&-[T_{\mu}^{~\mu}({\bf{r}}),[T_{\nu}^{~\nu}({\bf{r}'},-\tau),\rho_S(t)]]\Bigr)\cr
&&-iD({\bf{r}}-{\bf{r}'},\tau)\Bigl(2[T_{\mu\nu}({\bf{r}}),\{T^{\mu\nu}({\bf{r}'},-\tau),\rho_S(t)\}]\cr
&&-[T_{\mu}^{~\mu}({\bf{r}}),\{T_{\nu}^{~\nu}({\bf{r}'},-\tau),\rho_S(t)\}]\Bigr)\Bigr\},
\label{mastereq}
\end{eqnarray}
where $H_S$ is the free scalar field Hamiltonian and the noise and dissipation kernels are respectively: 
\begin{eqnarray}
N({\bf{r}},t)&=&\left(\frac{\kappa}{4}\right)^2\int\frac{d{\bf{k}}}{(2\pi)^3}\frac{e^{i{\bf k}\cdot {\bf r}}}{k} \cos(k t) [1+2n(k)]\cr
D({\bf{r}},t)&=&\left(\frac{\kappa}{4}\right)^2\int\frac{d{\bf{k}}}{(2\pi)^3}\frac{e^{i{\bf k}\cdot {\bf r}}}{k} \sin(k t),
\label{noiseeq}
\end{eqnarray}
with $n(k)$ the thermal Bose-Einstein occupation number at temperature $T$.

While the master equation~(\ref{mastereq}) can  in principle be used to investigate the  decoherence dynamics of quite general, relativistic scalar field matter states, we shall restrict ourselves to  scalar matter states that model ordinary, non-relativistic stationary macroscopic material objects. The following class of coherent states provides the basis for such a model:
\begin{equation}
|\alpha({\bf{k}})\rangle=\exp\left[-\frac{1}{2}\int d{\bf{k}}|\alpha({\bf{k}})|^2+\int d{\bf{k}}\alpha({\bf{k}})a^{\dagger}({\bf{k}})\right]|0\rangle,
\label{coherenteq}
\end{equation}
where 
\begin{equation}
\alpha({\bf{k}})=\varphi_0 R^3 \sqrt{\frac{\omega_m(k)}{2}}e^{-i{\bf{k}}\cdot{\bf{r}}_0-(kR)^2/2},
\label{alphaeq}
\end{equation}
with $\omega_m(k)=\sqrt{m^2+k^2}$. These states satisfy
\begin{eqnarray}
\langle\alpha({\bf{k}})|\phi({\bf{r}})|\alpha({\bf{k}})\rangle&=&\varphi_0 e^{-({\bf{r}}-{\bf{r}_0})^2/(2R^2)}\cr
\langle\alpha({\bf{k}})|\dot{\phi}({\bf{r}})|\alpha({\bf{k}})\rangle&=&0,
\label{expectationeq}
\end{eqnarray}
and thus describe Gaussian matter ``balls" of radius $R$ with stationary center at ${\bf{r}}_0$, and total energy content depending on the amplitude parameter $\varphi_0$. If we furthermore consider ball radii $R$ much larger than the nucleon's reduced Compton wavelength $\lambda_{\mathrm{C}}=\hbar/(mc)\approx 10^{-16}~{\mathrm{m}}$, then their rest mass energy $E=(\pi^3 m^2 \varphi_0^2 R^3)/2$ is the dominant energy content  and they approximately maintain their Gaussian profile (\ref{expectationeq}) with little spatial spreading over the timescale of the initial transient (see below); for simplicity we will neglect this spreading.       
The noise term part of the master equation~(\ref{mastereq}), which is responsible for decoherence, then simplifies to
\begin{eqnarray}
&&{\partial_t\rho_S[\phi,\phi',t]}=\cdots -\int_0^t d\tau\int d{\bf{r}}d{\bf{r}'}N({\bf{r}}-{\bf{r}'},\tau)\cr
&&\times\left[\frac{1}{2}m^2(\phi({\bf{r}}))^2-\frac{1}{2}m^2(\phi'({\bf{r}}))^2\right]\cr
&&\times\left[\frac{1}{2}m^2(\phi({\bf{r}'}))^2-\frac{1}{2}m^2(\phi'({\bf{r}'}))^2\right]\rho_S[\phi,\phi',t],
\label{masternoiseeq}
\end{eqnarray}
where we have used the fact that the energy density component  $T_{00}(\phi)\approx\frac{1}{2}m^2\phi^2$ of the energy-momentum tensor terms in Eq.~(\ref{mastereq}) dominates in the non-relativistic, stationary limit, and we have also expressed the master equation in the field coordinate basis.  

Let us now assume that, by some means, a superposition of two Gaussian ball states, each with distinct parameters $(\varphi_0,{\bf{r}}_0,R)$ and $(\tilde{\varphi}_0,{\tilde{\bf{r}}}_0,\tilde{R})$,  has been prepared at time $t=0$:
\begin{equation}
\rho_S[\phi,\phi',0]=\langle\phi|\Psi\rangle\langle\Psi|\phi'\rangle,
\label{initialstateeq}
\end{equation}
where
\begin{equation}
 \langle\phi|\Psi\rangle=\frac{1}{\sqrt{2}}\left(\langle\phi|\alpha({\bf{k}})\rangle+  \langle\phi|\tilde{\alpha}({\bf{k}})\rangle\right),
 \label{initialpurestateeq}
 \end{equation}
with the ball states in the field coordinate basis taking the form
\begin{eqnarray}
&&\langle\phi|\alpha({\bf{k}})\rangle\cr
&&=\exp\left[-\frac{1}{2}\int d{\bf{r}}\sqrt{m^2+ \nabla^2}\left(\phi({\bf{r}}) -\varphi_0 e^{-({\bf{r}}-{\bf{r}_0})^2/(2R^2)}\right)^2\right] \cr
&&\approx\exp\left[-\frac{m}{2}\int d{\bf{r}}\left(\phi({\bf{r}}) -\varphi_0 e^{-({\bf{r}}-{\bf{r}_0})^2/(2R^2)}\right)^2\right]
\label{coordinateballeq}
\end{eqnarray}  
and a similar expression for $\langle\phi|\tilde{\alpha}({\bf{k}})\rangle$ with parameters $(\tilde{\varphi}_0,{\tilde{\bf{r}}}_0,\tilde{R})$. The simpler approximate form in Eq.~(\ref{coordinateballeq}) follows from the condition $R\gg \lambda_{\mathrm{C}}$.  
Evaluating the noise term in~(\ref{masternoiseeq}) for the off-diagonal, interference part of the density matrix with $\phi({\bf{r}}) =\varphi_0 e^{-({\bf{r}}-{\bf{r}_0})^2/(2R^2)}$ and $\phi'({\bf{r}}) =\tilde{\varphi}_0 e^{-({\bf{r}}-{\tilde{\bf{r}}_0})^2/(2\tilde{R}^2)}$, we have
\begin{eqnarray}
&&{\partial_t\rho_S[\phi,\phi',t]}=\cdots \cr
&&-\frac{T}{2\pi}\left(\frac{\kappa}{4}\right)^2\left(\int d{\mathrm{r}}\left[\frac{1}{2}m^2(\phi({\bf{r}}))^2-\frac{1}{2}m^2(\phi'({\bf{r}}))^2\right]\right)^2\cr
&&\times\rho_S[\phi,\phi',t], 
\label{masternoise2eq}
\end{eqnarray}
where we neglect initial transients, corresponding to having $t$ large compared to the  time required for a graviton to traverse the matter state spatial extent, i.e., $ct\gg {\mathrm{max}}(\|{\bf{r}}_0-{\tilde{\bf{r}}_0}\|,R,\tilde{R})$--the Markovian approximation--and we also assume that $k_{\mathrm{B}}T\gg \hbar/t$--the high temperature limit. From Eq.~(\ref{masternoise2eq}), we immediately see that the off-diagonal interference part of the density matrix decays only provided the two ball states in the superposition have distinct energies $E\neq \tilde{E}$; spatial superpositions with ${{\bf{r}}_0}\neq{{\bf{r}}'_0}$ do not decohere if the respective energies are identical. Equation~(\ref{decohereeq}) immediately follows from~(\ref{masternoise2eq}). More precisely,  we have for the decoherence rate in the Born-Markov approximation:
\begin{equation}
\Gamma_{\mathrm{decohere}}=\frac{k_{\mathrm{B}} T}{\hbar}\left(\frac{E-\tilde{E}}{E_{\mathrm{P}}}\right)^2.
\label{decohere2eq}
\end{equation} 

\textit{Discussion}.--- The decoherence rate formula~(\ref{decohere2eq}) is sufficiently basic that one might expect it to be of more general validity beyond the specific scalar field model used above to derive it. Let us in particular assume that~(\ref{decohere2eq}) applies to ordinary, stationary  matter systems, such as a small chunk of crystalline solid or a trapped  cold atom cloud in the laboratory, and that for simplicity the matter system comprises model two state (excited and ground) atoms with energy level separation $\sim 1~{\mathrm{eV}}$.  For a cosmic gravitational wave background with temperature $T\sim 1~{\mathrm{K}}$~\cite{kolb90}, we have for the gravitationally induced decoherence rate of an initial  superposition of ground and excited states of a single atom: $\Gamma_{\mathrm{decohere}}\sim 10^{-45}~{\mathrm{secs}}^{-1}$. For a matter system comprising an Avogadro's number of  atoms  $\sim 1~{\mathrm{gram}}$ in a quantum superposition where all the atoms are either in their ground state or all in their excited state, then we have  $\Gamma_{\mathrm{decohere}}\sim 10^2~{\mathrm{sec}}^{-1}$. For a system with mass $\sim1~{\mathrm{kg}}$ in such a superposition state, the gravitationally induced decoherence rate is $\Gamma_{\mathrm{decohere}}\sim 10^8~{\mathrm{sec}}^{-1}$. Thus, even leaving aside the technical challenges due to the presence of everyday environments in preparing such macroscopic matter superposition states, the cosmic gravitational background itself will unavoidably induce their rapid decoherence, leaving the matter system in a classical mixture of either its ground or its excited state. In this way, we see that physical processes at the first instant of the Big Bang are ultimately responsible for enforcing classicality of ordinary macroscopic matter systems around today.

As  effective field calculations  go, the above $O(\kappa^2)$, Born-Markov derivation of the gravitationally induced decoherence rate is pretty straightforward; the present analysis should be viewed as a point of departure, showing the promise of the effective field theory approach~\cite{donoghue94,burgess04} for analyzing gravitationally induced decoherence. The calculations might be extended in several directions beyond the  master equation~(\ref{mastereq}), including (a)  going to $O(\kappa^4)$, so as to account for damping and decoherence due to graviton emission/absorption by the matter system; (b) investigating gravitationally induced decoherence for relativistic matter systems in curved space-time backgrounds, with application for example to the formation of cosmic matter structure in the early universe~\cite{perez06,unanue08}; (c) investigating the low temperature limit to determine whether gravity vacuum fluctuations can induce decoherence~\cite{diosi87,ellis89,penrose89,milburn91,blencowe91,power00,amelino00,gambini07,anastopoulos07,anastopoulos08,breuer09,gambini10}.

I would like to thank J. Ankerhold, R. R. Caldwell,  J. F. Donoghue, and R. Onofrio for very helpful discussions. This work was supported by the Carl Zeiss Foundation and  the National Science Foundation (NSF) under Grants No. DMR-0804477 and No. DMR-1104790.

\end{document}